\documentclass[preprint]{aastex}
\begin{document}
\newcommand{\msun}{M_{\odot}} \newcommand{\kms}{\, {\rm km\, s}^{-1}}
\newcommand{\cm}{\, {\rm cm}} \newcommand{\gm}{\, {\rm g}}
\newcommand{\erg}{\, {\rm erg}} \newcommand{\kpc}{\, {\rm kpc}}
\newcommand{\mpc}{\, {\rm Mpc}} \newcommand{\seg}{\, {\rm s}}
\newcommand{\kev}{\, {\rm keV}} \newcommand{\hz}{\, {\rm Hz}}
\newcommand{\etal}{et al.\ } \newcommand{\yr}{\, {\rm yr}}
\newcommand{\eq}{eq.\ } \newcommand{\lya}{Ly$\alpha$\ }
\newcommand{\hi}{\mbox{H\,{\scriptsize I}\ }}
\newcommand{\hii}{\mbox{H\,{\scriptsize II}\ }}
\newcommand{\hei}{\mbox{He\,{\scriptsize I}\ }}
\newcommand{\heii}{\mbox{He\,{\scriptsize II}\ }} \newcommand{\nhi}{N_{HI}}
\def\arcsec{''\hskip-3pt .}

\title{History of Hydrogen Reionization in the Cold Dark Matter Model}
\author{Christopher A. Onken \& Jordi Miralda-Escud\'e}
\affil{Department of Astronomy, The Ohio State University, Columbus, OH 43210}
\email{onken@astronomy.ohio-state.edu, jordi@astronomy.ohio-state.edu}

\begin{abstract}

  We calculate the reionization history in Cold Dark Matter (CDM)
models. The epoch of the end of reionization and the Thomson scattering
optical depth to the cosmic microwave background depend on the power
spectrum amplitude on small scales and on the ionizing photon emissivity per
unit mass in collapsed halos.  We calibrate the emissivity to reproduce the
measured ionizing background intensity at $z=4$. Models in which all CDM
halos have either a constant emissivity or a constant energy emitted per
Hubble time, per unit mass, predict that reionization ends near $z\sim 6$
and the optical depth is in the range $0.05 < \tau_e < 0.09$, consistent
with {\it WMAP} results at the $1\sigma$ to $2\sigma$ level. If the optical
depth is as high as $0.17$ (as suggested by {\it WMAP}), halos of velocity
dispersion $\sim 3 - 30 \kms$ at $z>15$ must have ionizing emissivities per
unit mass larger by a factor $\gtrsim 50$ compared to the more massive halos
that produce the ionizing emissivity at $z=4$. This factor increases to 100
if the CDM power spectrum amplitude is required to agree with the Croft
\etal (2002) measurement from the \lya forest. If $\tau_e \gtrsim 0.17$ were
confirmed, a higher ionizing emissivity at $z>15$ compared to $z=4$ might
arise from an enhanced star formation rate or quasar abundance per unit mass
and an increased escape fraction for ionizing photons; the end of
reionization could have been delayed to $z\sim 6$ because of the suppression
of gas accretion and star formation in low-mass halos as the medium was
reionized.

\end{abstract}


\section{INTRODUCTION}

  The epoch of reionization of the universe started with the emission of the
first ionizing photons into the intergalactic medium (hereafter, IGM), and
ended when all the low-density regions of the universe were ionized. During
this period of time, the ionizing photons reaching the IGM had to be
sufficient to ionize every atom in the universe and to balance any
recombinations in the IGM (in addition, many more ionizing photons may have
been emitted in regions of dense, self-shielded gas, which were locally
absorbed and did not contribute to ionizing the IGM).  At the end of
reionization, the mean free path of ionizing photons increased to scales
much larger than the typical separation of the collapsed halos hosting the
sources. Regions of dense, self-shielded gas (observed as Lyman limit
systems) shrank in size as the ionizing background increased in intensity,
due to the growing mean free path (Miralda-Escud\'e, Haehnelt, \& Rees
2000). If the fraction of baryons in regions of intermediate density which
were ionized as the mean free path increased was small, the rise in the
background intensity could be relatively fast, as seen in the numerical
simulations of Gnedin (2000).

  Precisely this rapid change of the ionizing background intensity has been
inferred by Fan \etal (2002) to occur at $z\simeq 6$, by analyzing the
change of the mean transmitted flux of the \lya forest in the highest
redshift quasars known. The transmitted flux apparently drops abruptly to
very low levels near $z=6$. This observation most likely implies that the
end of reionization occurred, in fact, at $z\simeq 6$, although as a larger
number of quasars at $z>6$ are discovered, more detailed investigations of
this issue will need to be done to understand precisely how the mean free
path grew. In particular, the number of small gaps of transmitted flux, such
as the one seen in a $z=6.37$ quasar by White \etal (2003), provides
information on the size of the \hii regions before they overlapped
(Miralda-Escud\'e 1998; Barkana 2002).

  The results of the {\it Wilkinson Microwave Anisotropy Probe} ({\it WMAP};
Bennett \etal 2003 and references therein) have strengthened the
observational support of the Cold Dark Matter (CDM) scenario in a flat model
that contains a vacuum energy component (in addition to the ordinary
baryonic matter and dark matter).  The model parameters are being measured
with increasing accuracy via combined observations of the cosmic microwave
background (CMB), galaxy clustering, weak lensing, and the \lya forest. The
theory then makes an increasingly reliable prediction for the abundance of
collapsed dark matter halos as a function of mass and redshift. It is in
these halos that the process of radiative gas dissipation and star formation
can take place, leading to the emission of ionizing radiation. However, the
uncertainties in the efficiency of star formation, the stellar initial mass
function (IMF), the emission due to gas accretion into black holes (be they
in X-ray binaries or active galactic nuclei), and the fraction of the
ionizing photons that escape into the IGM from the dense regions in which
they are emitted, prevent a clear prediction for the history of
reionization.

  Even if the time at which reionization {\it ends} is determined to be at
$z\simeq 6$ by the Gunn-Peterson trough (Gunn \& Peterson 1965)
observations, there can be a long time interval in which the universe was
partially ionized, with cosmological \hii regions around the sources filling
only some fraction of the volume. A very good observational probe for the
characteristic time at which the bulk of the baryons in the universe were
ionized is the electron scattering optical depth to the CMB, $\tau_e$. This
has been measured for the first time by {\it WMAP} (Kogut \etal 2003;
Spergel \etal 2003) using the large-scale polarization-temperature
anisotropy correlation. At present, the result still has a large error, both
observational and model-dependent: Kogut \etal find a model-independent
value of $\tau_e=0.16 \pm 0.04$ from the polarization-temperature
correlation of the {\it WMAP} data, but Spergel \etal find $\tau_e=0.17 \pm
0.06$ from a fit of a CDM model with a running spectral index to a combined
data set of CMB, galaxy clustering and \lya forest observations, and
$\tau_e=0.12 \pm 0.06$ when the fit is forced to the usual CDM model with a
primordial power-law spectrum. A value of $\tau_e=0.17$ would imply that
most baryons were reionized as early as $z\sim 20$.

  The announcement of the measurement of $\tau_e$ by {\it WMAP} has led to a
large number of papers discussing the possibility of this early reionization
in the CDM model (Whyithe \& Loeb 2003; Haiman \& Holder 2003; Holder \etal
2003; Ciardi, Ferrara, \& White 2003; Somerville \& Livio 2003; Sokasian
\etal 2003; Cen 2003; Chiu, Fan, \& Ostriker 2003). These authors have
generally found that, if one is willing to assume a high efficiency in the
production of ionizing photons per unit mass in the low-mass halos
collapsing at early times, then early reionization and a high optical depth
is possible in CDM models.  Naturally, the lower the power spectrum
amplitude on small scales, the higher the efficiency that is required, and
some models can be ruled out based on their small power on small scales,
such as the warm dark matter model (Spergel \etal 2003). But to a large
extent, the ability of the CDM model to accommodate a wide range for the
epoch at which the bulk of the IGM was reionized simply reflects our
ignorance of the efficiency in producing ionizing photons that escape.

  However, a reasonable question we can ask is how the required efficiency
compares to any observationally determined values of the rate at which
ionizing photons have been emitted over the history of the universe, taking
into account the predicted abundances of halos of different masses that form
at each redshift. In this paper, we adopt a very simple model of the
ionizing emissivity from CDM halos, but which still captures the essential
physical ingredient for the process of reionization: as the medium is
ionized, the gas is heated and the Jeans mass is raised, implying that some
low-mass halos where gas can collapse when the medium is neutral are unable
to continue to accrete gas after the medium is ionized. We divide halos into
two populations, the ``high-mass'' ones (which will be called Population A)
that are able to accrete gas and form new stars irrespective of the state of
the intergalactic medium, and the ``low-mass'' ones (Population B) that can
form stars and emit ionizing radiation only when the medium is neutral.
Low-mass halos collapse first in CDM, but as they ionize the medium around
them the rate of star formation can be suppressed, until the high-mass halos
start to collapse (Couchman \& Rees 1986; Shapiro, Giroux, \& Babul 1994). A
similar separation of halo classes was made by Haiman \& Holder (2003); our
Population A corresponds to their Type II and Type Ia halos (Haiman \&
Holder separated these halos into the ones that can or cannot cool by atomic
cooling after they have accreted the gas; we will not distinguish between
these two classes in our more simple model), and our Population B
corresponds to their Type Ib population.

  Our calibration of the efficiency to produce ionizing photons is based on
observations of the \lya forest during the post-reionization era. The mean
transmitted flux of the \lya forest, combined with CDM simulations of \lya
forest spectra and measurements of the baryon density of the universe,
allows us to infer the intensity of the ionizing background; and the
abundance of Lyman limit systems tells us the mean free path of ionizing
photons. The ratio of the background intensity to the mean free path then
yields the volume-averaged emissivity. This emissivity was inferred to be,
at most, seven ionizing photons per baryon and per Hubble time at $z=4$ by
Miralda-Escud\'e (2003). Remaining uncertainties on the value of this
emissivity are related mainly to the temperature of the IGM and the CDM
power spectrum amplitude. At $z>4$, the abundance of Lyman limit systems
becomes highly uncertain, so the emissivity cannot be inferred. The key
question for reionization is, of course, how this emissivity changes with
increasing redshift. If the comoving emissivity stays constant at $z>4$, the
optical depth $\tau_e$ is not larger than $0.09$ (Miralda-Escud\'e 2003). In
CDM models, the fraction of mass in halos that can form stars decreases with
redshift at $z>4$, so to obtain $\tau_e > 0.09$ (which requires an increase
of the emissivity with redshift), the low-mass halos (or Population B)
formed at $z \gg 4$ must have a much higher efficiency to emit ionizing
radiation than the halos of higher mass at $z\simeq 4$.

  The question we are addressing in this paper is: how much higher does this
efficiency have to be in order to produce a certain value of the optical
depth $\tau_e$, and what constraints are imposed by requiring reionization
to end at $z=6$? We describe our model for the efficiency to emit ionizing
radiation in \S 2. Results are presented in \S 3, and a general discussion
is given in \S 4.

\section{MODELS}

\subsection{CDM Power Spectrum}

  We use three CDM models to predict the halo abundances. All the models
assume flat space with a cosmological constant, with $\Omega_{m0}=0.27$,
$\Omega_{b0}=0.044$, $h=0.71$. The three models vary in the normalization
and slope of the power spectrum: Model 1 has $\sigma_8=0.84$, $n_s=0.93$;
Model 2 has $\sigma_8=0.9$, $n_s=0.96$; and Model 3 has $\sigma_8=0.9$,
$n_s=1$ (here, $\sigma_8$ is the linear rms mass fluctuation at present on
spheres of radius $8 h^{-1} \mpc$, and $n_s$ is the primordial scalar
spectral index). The power spectrum is computed with the fitting formula of
Eisenstein \& Hu (1999). Model 1 has the smallest amplitude of fluctuations
on the small scales on which the first halos that emit ionizing radiation
collapse ($0.03$ to 1 Mpc), and Model 3 has the largest amplitude. Model 3,
with the largest power on small scales, has halos forming at the highest
redshift, and therefore predicts an earlier start of reionization and higher
optical depth $\tau_e$ than the other models for a fixed ionizing emissivity
per unit mass.

  Model 1 has the same parameters as the best fit model found by {\it WMAP}
(Bennett \etal 2003; Spergel \etal 2003) to the combined data set of CMB
observations, the 2dF Galaxy Redshift Survey (Colless \etal 2001), and \lya
forest power spectrum data (Croft \etal 2002). However, we do not include
the running spectral index introduced in Spergel \etal (inclusion of the
running spectral index would reduce the power on small scales). Model 2 has
higher values of $\sigma_8$ and $n_s$ by $1.5$ and $1$ standard deviations,
respectively, from the best fit values of Spergel et al. Model 2 also has a
slightly larger power than the best fit CDM model without a running spectral
index found by Spergel \etal (which has parameters $\sigma_8=0.8$ and
$n_s=0.96$). Finally, Model 3 is very close to the best fit of Spergel \etal
to CMB data alone, and is the same model that was adopted in various recent
studies on the reionization history (Ciardi \etal 2003; Somerville \& Livio
2003; Sokasian \etal 2003; Cen 2003).

  Measurement of the \lya forest has allowed a determination of the power
spectrum amplitude on the smallest scales. We therefore compare the
amplitude of our models with the determination of the \lya forest power
spectrum by Croft \etal (2002), who find a normalization at redshift
$z=2.72$ of
\begin{equation}
\Delta^2(k_p) \equiv {k_p^3\over 2\pi^2 }\, P(k_p) =
0.74^{+0.20}_{-0.16} ~,
\end{equation}
where $H(z)^{-1}\, k_p=0.03 (\rm km/s)^{-1}$. The normalization of our
Models (1, 2, 3) are $\Delta^2(k_p)=(0.85, 1.06, 1.27)$, respectively.
Hence, Models 2 and 3 have a power spectrum amplitude higher than the Croft
\etal (2002) measurement by $1.6\sigma$ and $2.6\sigma$, respectively. The
high amplitude of Model 3 would seem to be ruled out by the result of Croft
et al.; note, however, that the quoted error on this measurement does not
include some potentially important systematic effects, as discussed in Croft
et al., and the amplitude can be higher if the mean \lya transmitted flux is
higher than the value used by Croft \etal (see Seljak, McDonald, \& Makarov
2003).

\subsection{Halo Abundances}

  We use the Sheth \& Tormen (1999) prescription for computing the
abundances of halos as a function of mass and redshift. Halos are divided
into two populations: those which do not emit any ionizing photons after the
medium around them has been reionized (Population B), and those which are
massive enough to continue forming stars and producing ionizing photons
after reionization (Population A). In this paper we place the division
between the two populations at a velocity dispersion $\sigma_{0}=35 \kms$,
based on the suppression of gas infall and cooling in smaller halos caused
by photoionization (Efstathiou 1992; Thoul \& Weinberg 1996; Navarro \&
Steinmetz 1997).  The lower limit to the velocity dispersion of the
Population B of low-mass halos is placed at $\sigma_{min}=3.68 \kms$,
because molecular hydrogen cooling below a temperature of $\sim 2000$ K is
ineffective (Yoshida \etal 2003).

\begin{figure}
\plotone{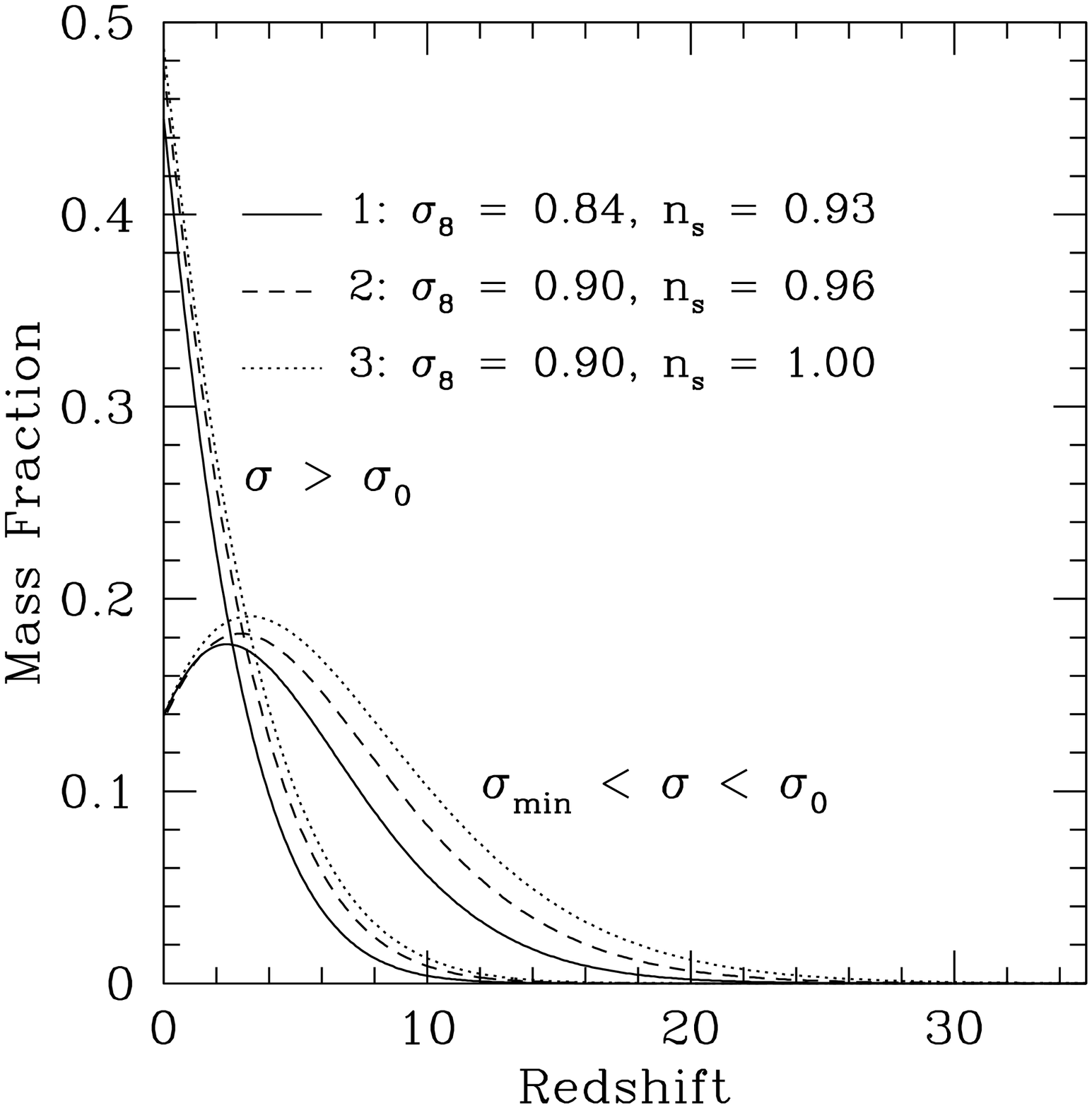}
\caption[f1.eps]{Fraction of mass collapsed into halos as a function of
redshift for two ranges in halo velocity dispersion, $\sigma$. Population A
includes the halos with $\sigma > \sigma_{0}$, while Population B represents
the halos with $\sigma_{min} < \sigma < \sigma_{0}$. For each of the Model 1
(solid), Model 2 (dashed), and Model 3 (dotted) cosmological parameter sets,
$\sigma_{min}=3.68 \kms$ and $\sigma_{0}=35 \kms$. \label{fig1}}
\end{figure}

\begin{figure}
\plotone{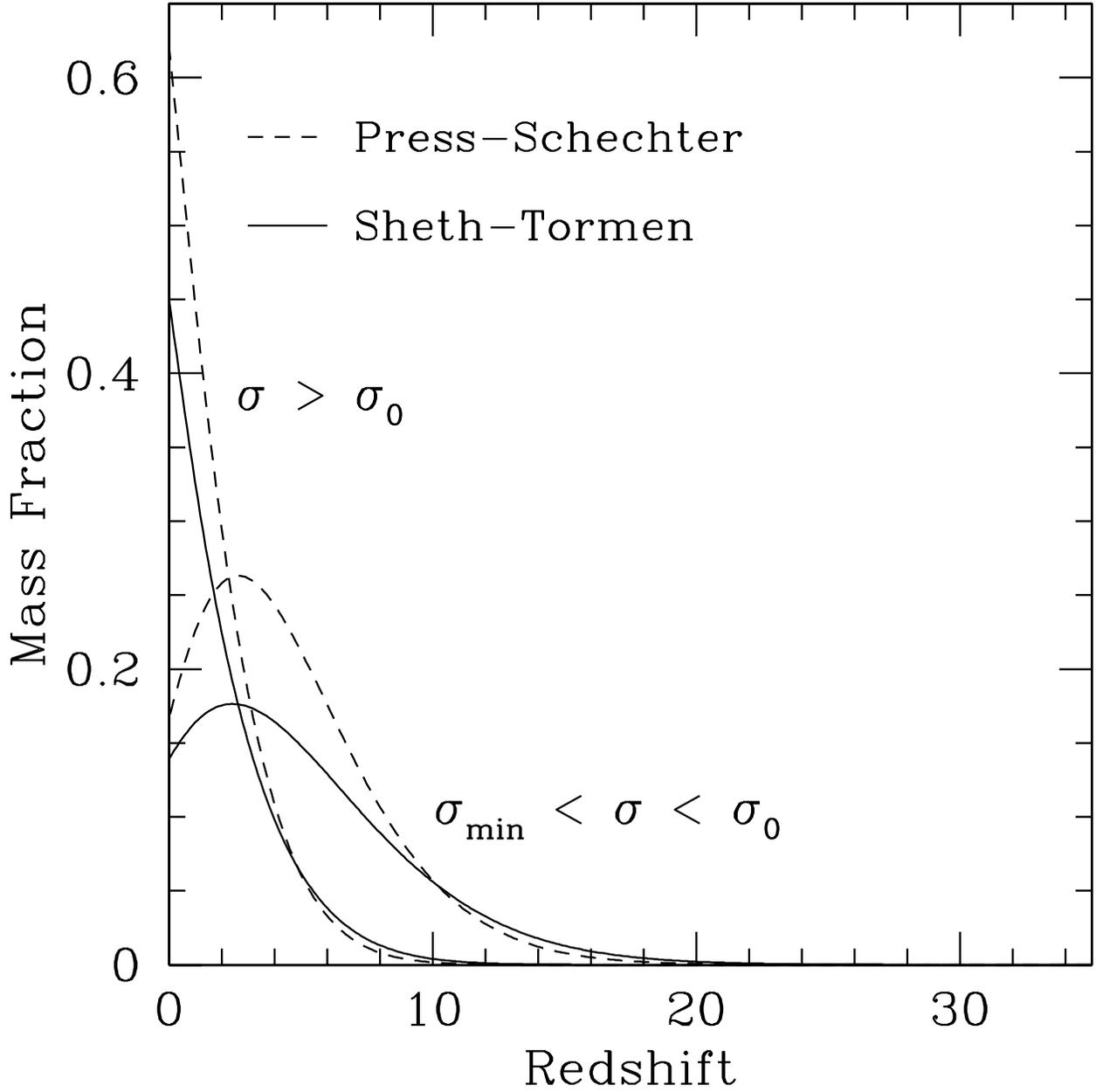}
\caption[f2.eps]{Comparison of the amount of collapsed mass in halo
Populations A and B as a function of redshift derived with the Sheth-Tormen
(solid) and Press-Schechter (dashed) mass functions for Model
1. \label{fig2}}
\end{figure}

  The fraction of baryons, $F_{A,B}$, which occupy each of the halo
populations is shown in Figure \ref{fig1} for Models 1 to 3. We also show in
Figure \ref{fig2} the difference between using the Sheth-Tormen and the
Press-Schechter (Press \& Schechter 1974) mass functions, for Model 1.  The
Sheth-Tormen mass function predicts a slightly larger number of objects at
the highest redshifts.

\subsection{Emissivity}

  For Population A, i.e., halos with $\sigma > \sigma_{0}=35 \kms$, the
ionizing photon emissivity per baryon is assumed to be of the form
\begin{equation}
\epsilon_A(z) = F_A(z) \frac{\epsilon_{4}}{F_A(4)}
\left({1+z \over 5}\right)^{\alpha} ~,
\label{emisa}
\end{equation}
where $F_A(z)$ is the fraction of baryons in Population A halos at redshift
$z$, and $\epsilon_4$ is the total emissivity at $z=4$, which is assumed to
be equal to 7 ionizing photons per baryon and per Hubble time at $z=4$, the
upper limit obtained in Miralda-Escud\'e (2003). We will use two values of
$\alpha$ in this paper: for $\alpha=0$ the energy emitted per unit physical
time and per unit mass remains constant, and for $\alpha=1.5$ the total
energy emitted per Hubble time at redshift $z$ remains constant. Note that
gas cooling is inefficient in very massive halos when the virialized gas
temperature is $> 10^7 K$, but at $z=4$ the abundance of these halos, which
arise from $>4\sigma$ fluctuations, are negligibly small, so our assumption
of a constant emissivity per unit mass for Population A is reasonable.

  For Population B, we allow an emissivity per unit mass different from the
one in Population A halos, parameterizing it as
\begin{equation}
\epsilon(\sigma) = 
\left[ 1 + \gamma\ln \frac{\sigma_{0}}{\sigma} \right]
\frac{\epsilon_{4}}{F_A(4)} \left({1+z \over 5}\right)^{\alpha} ~,
\label{emisb}
\end{equation}
where $\gamma$ is a scaling factor used to adjust the variation in
efficiency between the low-mass and high-mass halos. This results in a
smooth variation of the emissivity with $\sigma$.  The Population B
emissivity is then
\begin{equation}
\epsilon_{B}(z) = \int_{\sigma_{min}}^{\sigma_{0}} f_B(\sigma,z)\,
\epsilon(\sigma)\, d\sigma ~,
\end{equation}
where $f_B$ is the fraction of mass in halos in the interval $d\sigma$
around $\sigma$, and we set $\sigma_{min}=3.68\kms$, corresponding to $2000$
K for the mean particle mass of the primordial atomic gas.

\subsection{Analytical Framework}

The equation for the ionization of the intergalactic medium is (Madau,
Haardt, \& Rees 1999; Miralda-Escud\'e 2003)
\begin{eqnarray} 
\frac{dy}{dt} = & \epsilon - R y \\
 = & \epsilon_A + \epsilon_B (1 - y) - R y, \label{eq5}
\end{eqnarray}
where $y$ is the volume fraction of the IGM that is ionized at any given
time $t$, and $R$ is the mean number of recombinations per baryon in the
ionized regions per unit time. This equation makes the approximation that
the photon mean free path is very short, so emitted photons are
instantaneously used to ionize atoms. It also hides all the complicating
physics of reionization in the averaged recombination rate $R$, which
depends on the clumping factor of the photoionized gas. Note that Population
B halos emit ionizing radiation only when they are located in the atomic
medium, so the emissivity $\epsilon_B$ is multiplied by $1-y$. We assume a
clumping factor of unity throughout this paper, computing the recombination
coefficient at $T=10^4$ K, which gives $R=1.16\times10^{-17}\, {\rm s}^{-1}$
at $z=4$.  The solution of equation (\ref{eq5}) is
\begin{eqnarray}
y(t) = &\int_{t_i}^t\, \exp \left\{ - {\int_{t'}^{t}
\left[ R(t'') + \epsilon_B(t'') \right] dt'' }\right\} \nonumber \\
       &\times \left[ \epsilon_B(t') + \epsilon_A(t') \right]\, dt' ~,
\end{eqnarray}
where $t_i$ is an arbitrarily chosen initial time (before any appreciable
halo formation).

\subsection{Optical Depth}

  The Thomson scattering optical depth, $\tau_e$, is
\begin{equation}
\tau_e(z) = \frac{n_{e} \sigma_{e} c}{H_{0}} \int_{0}^{z} 
\frac{F_{i} (1+z)^{2}\, y(z)\, dz}{[\Omega_{\Lambda0} + \Omega_{m0}
(1+z)^{3}]^{1/2}} ~,
\end{equation}
where $\Omega_{\Lambda 0} = 1- \Omega_{m0}$, $n_e$ is the comoving electron
number density, $\sigma_e$ is the electron cross section, and $F_i$ is the
fraction of the baryons in the ionized regions that are actually in the
ionized IGM (with the rest being in self-shielded atomic or molecular
clouds, or in stars). We assume $F_i=0.9$ (as in Miralda-Escud\'e 2003) for
$z<6$, but $F_i=1$ at $z>6$; lowering $F_i$ at $z>6$ would demand starting
reionization at higher redshift to obtain a fixed optical depth, increasing
the required emissivity from the first star-forming halos (note that
equation (5) needs to be modified if $F_i < 1$ during reionization). For our
value of $\Omega_b h^2=0.022$, the electron number density is $n_e=2.7\times
10^{-7} (1 - f_Y Y) \cm^{-3}$, where $Y=0.24$ is the helium abundance by
mass, and $f_Y=(0.5, 0.75)$ when the helium is (doubly, singly) ionized. We
assume that helium becomes doubly ionized at $z=3.5$ (e.g., Heap \etal
2000), and that it is singly ionized at higher redshift wherever hydrogen is
ionized. This results in an optical depth up to $z=6$ of
$\tau_e(z=6)=0.032$.

\section{RESULTS}

  We now present the results on the fraction of the volume that is ionized
as a function of redshift, $y(z)$, with the emissivity that arises from our
two populations of halos given by equations (\ref{emisa}) and
(\ref{emisb}). To summarize the emissivity model, the Population A
(high-mass halos) emissivity per unit mass is fixed at $z=4$ and is allowed
to vary as $(1+z)^{\alpha}$, while the Population B (low-mass halo)
emissivity per unit mass can also be increased with the parameter $\gamma$.

 Figure 3 shows $y(z)$ for each of the three CDM models and for both $\alpha
= 0$ and $\alpha=1.5$.  The case $\gamma=0$, where the emissivity per unit
mass from Population A and B is the same, is shown as solid lines. The cases
$\gamma=1$, $10$ and $100$ are shown as dotted, dashed, and dash-dot lines,
respectively. The models with high $\gamma$ values have an early
reionization as the first Population B halos collapse; then, $y(z)$ grows
slowly when it approaches unity because of the decrease in the emission from
Population B as most of the IGM is ionized, and finally $y$ reaches unity
when the Population A halos start forming and emitting radiation in the
ionized regions of the universe.

  The optical depth to electron scattering for all the models shown is
listed in Table 1.

  The emissivity as a function of redshift is shown in Figure 4 for the same
models. At redshift just above $4$, the emissivity declines with redshift
because the abundance of Population A halos (which are the only ones
contributing when the medium is entirely ionized) drops rapidly.  The
emissivity increases again at redshifts higher than the end of reionization,
when Population B halos contribute. Models with higher $\gamma$ can
obviously have a higher emissivity at very high redshift, but their
emissivity declines at intermediate redshifts when $y$ is close to unity and
the emission from Population B halos is suppressed, as assumed in equation
(\ref{emisb}). The models with higher $\gamma$ result in higher optical
depths because the IGM is reionized earlier.

\section{DISCUSSION}

  For a fixed CDM halo population, the most simple model for the emissivity
is to say that the rate of photon emission per unit mass is constant for all
halos in which cooling can take place, i.e., with $\sigma >
\sigma_{min}$. This model, using the emissivity per unit mass calibrated
with the observational determination at $z=4$ and a clumping factor of one,
predicts a redshift at which reionization ends of $5$ to $7$, in very good
agreement with the observation of the appearance of the Gunn-Peterson trough
at $z=6$. The emissivity history in this model shows a decline with redshift
from $z=4$ to $6$, then a slight increase up to $z\sim 10$ as more
Population B halos contribute, and then a rapid decline beyond $z\sim$ 10 to
15 as the number of all halos decreases.

  Obviously, the reionization history should also depend on the clumping
factor affecting the recombination rate. We have assumed a clumping factor
of unity, and a higher clumping factor would delay the end of
reionization. A moderate clumping factor of 2 or 3 would not greatly affect
the result, because the recombination time at the mean baryonic density is
equal to the Hubble time at $z=6$ and the increase in $y(z)$ is relatively
fast for the $\gamma=0$ models (solid lines in Fig.\ 3; this is simply due
to the rapid increase in the mass fraction in collapsed halos shown in Fig.\
1), so the number of recombinations is not very large. For example, a
clumping factor of 3 for Model 1 with $\alpha=1.5$ (panel b) delays the end
of reionization from $z=6$ to $z=5$. As discussed by Miralda-Escud\'e \etal
(2000), and Miralda-Escud\'e (2003), the clumping factor is unlikely to be
large during reionization. Therefore, we have the remarkable result that the
simplest model of a constant emissivity per unit mass in CDM halos,
calibrated with the observed emissivity at $z=4$, roughly predicts the right
epoch for the end of reionization.

  This same model predicts an optical depth to Thomson scattering, $\tau_e$,
of $0.05$ to $0.09$, depending on $\alpha$ and on the allowed variations on
the power spectrum parameters determining the amplitude of fluctuations on
the small scales of the first halos. Although this is smaller than the value
found by the {\it WMAP} mission in a CDM model fit according to Spergel
\etal (2003), $\tau_e=0.17 \pm 0.06$, it is not ruled out with the present
errorbar: the upper value of $\tau_e=0.09$ of our $\gamma=0$ models is
consistent with this measurement at the $1.3\sigma$ level. The likelihood
function of this range of $\tau_e$ shown in Figure 8 of Spergel \etal (2003)
is also seen to be not very small. Moreover, if the running spectral index
in the CDM model is not allowed, the best fit then yields $\tau_e=0.12 \pm
0.06$ (see Table 7 of Spergel et al.). The smaller errorbar obtained by the
analysis in Kogut \etal (2003) is puzzling, because the addition of
observational data and the use of a parameterized model instead of a
model-independent analysis should not result in an increased error.

  If a high value of $\tau_e$ is confirmed, then the ionizing emission
efficiency in the earliest low-mass halos to form in CDM is required to be
much higher than in the sources present in the more massive halos at
$z=4$. The required factor by which the efficiency must increase can be
inferred from Table 1. The increase in emissivity per unit mass is $\sim
(1+2\gamma)(1+z)^{\alpha}$ (see eq.\ 3), for a typical halo contributing to
the emissivity at high redshift with $\sigma\sim 5 \kms$. For the case
$\alpha=0$, the factor $1+2\gamma$ gives the increase in the {\it rate} of
emission per unit mass, while for the case $\alpha=1.5$, the factor
$1+2\gamma$ gives the increase in the {\it total} energy per unit mass
emitted in ionizing photons, taking into account the shorter Hubble time at
higher redshift. Interpolating from Table 1, we find that to reach
$\tau_e=0.17$, the total ionizing energy per unit mass emitted by Population
B halos at $z\gtrsim 15$ needs to be larger by a factor $1+ 2\gamma\sim 10$
compared to the Population A halos at $z=4$, and the rate of emission needs
to be higher by a factor $\sim 100$. For Model 1 (more consistent with the
Croft \etal 2002 measurement of the \lya forest power spectrum), the energy
emitted per unit mass needs to increase by a factor $\sim 25$, and the
emission rate per unit mass needs to increase by $\sim 300$.

  Are these very high efficiencies for producing ionizing photons in the
earliest low-mass halos physically realistic? The emissivity of Population A
halos at $z=4$ is about 60 photons per baryon per Hubble time, as obtained
by dividing the mean emissivity $\epsilon_4$ of 7 photons per baryon per
Hubble time (see \S 2.3) by the fraction of baryons in Population A halos at
$z=4$ (see Fig.\ 1). From the emission ratios mentioned above required to
produce $\tau_e=0.17$, Population B halos at $z \simeq 20$ would need to
emit $\sim 1500$ and $600$ photons per Hubble time for Models 1, and 3,
respectively. Since in these models with high optical depth the universe is
reionized at $z\simeq 20$ over a time interval corresponding to $\Delta z
\simeq 5$ (see Fig.\ 3), or one quarter of a Hubble time, the total emission
during this time interval coming from the first halos needs to be about
$400$ and $150$ photons per baryon in Population B halos, for Models 1 and
3.

  To compare this to the number of photons that can be produced by star
formation, we use the models of Leitherer \etal (1999; see their Fig.\ 78)
to find that for a Salpeter IMF with solar metallicity in the stellar mass
range $1 \msun < M < 100 \msun$, 6000 ionizing photons per baryon are
emitted. Most of the ionizing photons are produced by stars with $M> 20
\msun$, and in the case of a Salpeter IMF with stars forming only in the
range $20 \msun < M < 100 \msun$, 25000 photons per baryon can be
emitted. Note that this number is equal to the ratio of the energy obtained
by hydrogen fusion (7 MeV) to the average energy of an ionizing photon (20
eV) times the fraction of stellar mass that is fused to helium ($\sim 30$
\%) times the fraction of energy emitted as ionizing photons ($\sim
25$\%). Very massive stars with zero metallicity can further increase this
emission to $10^5$ photons per baryon (Tumlinson \& Shull 2000; Bromm,
Kudritzki, \& Loeb 2001; Schaerer 2002), because their convective cores
during the main-sequence include almost all the mass of the star (zero
metallicity stars also have higher temperature than metal-rich ones during
the main-sequence, which increases the fraction of energy they emit in
ionizing radiation, but also increases the average energy of their ionizing
photons). Thus, if $f_*$ is the fraction of the gas in a halo that can be
converted to stars with $M> 20 \msun$, and $f_{esc}$ is the fraction of
ionizing photons that escape to the IGM, then an optical depth $\tau_e=0.17$
requires $f_* f_{esc} > (10^{-2.4}, 10^{-2.8})$ for Models (1,3) for
zero-metallicity stars, and $f_* f_{esc} > (10^{-1.8}, 10^{-2.2})$ for
Models (1,3) for metal-enriched stars.

  It is clear from these values that an optical depth $\tau_e=0.17$ can be
produced with physically realistic models. Even $\tau_e=0.3$ is possible in
Model 3 with $\gamma=100$ and $\alpha=1.5$, which implies an emission of
3000 ionizing photons per baryon (or $f_* f_{esc}=10^{-1.5}$ for
zero-metallicity stars) from the first star-forming halos over the redshift
range $25 \lesssim z \lesssim 32$. However, if a high optical depth is
confirmed by future measurements, an explanation for the dramatic change in
ionizing emission from the low-mass halos that formed the first stars to the
more massive halos at lower redshift will need to be found.

  A difficulty in making star formation very efficient in low-mass halos is
that their shallow potential wells makes them more vulnerable to lose their
gas as a result of stellar winds, photoionization and supernovae.  The
escape fraction might be higher than in more massive halos at lower
redshift, but only after gas has been ionized and expelled from the vicinity
of stars, in which case the gas can easily escape from the halo. The typical
halo where stars form at $z\simeq 20$ with velocity dispersion $\sim 5 \kms$
and total mass of $10^6 \msun$ (e.g., Yoshida \etal 2003) has a gas mass of
$\sim 2\times 10^5 \msun$ within a virialized radius $\sim 200$ parsecs, and
cooling by molecular hydrogen leads to the formation of a dense central gas
clump containing only $\sim$ 1\% of the gas mass. The rest of the gas must
take more than a halo dynamical time ($\sim 10^{7.5}$ years) to reach the
center and cool before it has any possibility to form stars, by which time
supernovae would have taken place.  The hydrodynamic simulations of these
first star-forming halos (Abel, Bryan, \& Norman 2002; Bromm, Coppi, \&
Larson 2002) suggest that the first stars may all be very massive. If the
$\sim 2000 \msun$ of the core of cooled gas can efficiently form one or a
few massive stars, and most of the emitted ionizing photons escape, one may
barely be able to achieve the fraction $f_* f_{esc}\sim 10^{-2.5}$ required
for $\tau_e=0.17$. After the formation of these stars, the supernovae
explosions would expel the remaining gas from the halo (Bromm, Yoshida, \&
Hernquist 2003).

  In halos of higher mass (with $\sigma$ of 10 to 30 $\kms$), a larger
number of supernovae would be required to expel the gas. Nevertheless, the
effects of photoionization and supernova explosions should still
self-regulate the rate of massive star formation, and it is not clear why
the earliest galaxies at high redshift could have higher values of $f_*
f_{esc}$ than the more massive galaxies formed at lower redshift.  While the
mean gas density in galaxies should increase with redshift, increasing gas
cooling rates, the shallower potential wells of low-mass halos allow for gas
dispersal after a smaller fraction $f_*$ of gas has turned to massive stars.

  Another possibility is that black holes formed by the first massive stars
are able to accrete gas. Gravitational accretion can produce radiation with
an efficiency $\sim 20$ times greater than stellar fusion (and stars
generally fuse only a fraction of their mass). In this case, requiring a
fraction as low as $\sim 10^{-3.5}$ of the halo gas to accrete on a central
black hole could produce the $\sim 300$ photons per baryon in halos required
for $\tau_e=0.17$, depending on the fraction of the accretion energy emitted
as ionizing photons (we have assumed 20\% here).  However, whether one
speculates about a high efficiency of massive star formation or of gas
accretion into black holes, the question that remains unanswered is why such
high efficiency to produce ionizing radiation would occur only in the first
halos at high redshift, and not in similar or more massive halos at later
times.

  If it were nevertheless possible to produce these high emissivities when
the first stars formed, a second possible problem of an early reionization
might be that the high emissivity leads to a rapid end of reionization, in
contradiction with the presence of the Gunn-Peterson trough at $z=6$. Our
results suggest a way to solve this particular problem: as seen in Figure 3,
once the fraction of the IGM that is ionized becomes close to unity, the
emissivity can be decreased because star formation in Population B halos is
suppressed. This can regulate the rate of emission to a level that just
balances recombinations, and it is only when the more massive Population A
halos reach a critical abundance that the emissivity can rise again,
completing reionization and causing the fast increase in the mean free path
of ionizing photons that clears the Gunn-Peterson trough. Thus, in all our
models the end of reionization occurs between redshift 5 and 8, even though
the models with high $\gamma$ can ionize half of all the volume as early as
$z \simeq 20$. In the models with the earliest reionization in Figure 3, the
redshift for the end of reionization could easily be lowered to fit the
observed value of 6 by moderately increasing the clumping factor after $y$
is close to unity or increasing the value of $\sigma_0$ that separates our
Populations A and B.

  Note that models with high $\gamma$ have an emissivity history that is
double peaked: there is a first maximum due to Population B halos, then the
emissivity declines as most of the IGM is ionized, and a second maximum is
then caused by Population A halos. The possibility of a double-peaked
history of star formation due to the effect of reionization raising the
Jeans mass was proposed by Couchman \& Rees (1986) and Shapiro et
al. (1994). However, this does not necessarily imply a ``double
reionization'', as other authors have argued (Cen 2003; Whyithe \& Loeb
2003). The fraction of ionized volume $y(z)$ rises monotonically with time,
and it is only ``stalled'' by the suppression of the star formation due to
ionization, in a self-regulated process. A double reionization in which the
universe recombines again after having been ionized would require a long
delay of the feedback ionization effects on the star formation rate. If this
delay were present, the suppression of star formation might start to operate
only after stars have already emitted more than enough photons to ionize all
the universe; later, when the IGM recombines again, star formation might not
resume at the same level for some time, allowing for partial recombination
to occur. This feedback delay might be caused by a reservoir of gas in
Population B halos that can keep forming stars after the surrounding
low-density medium is ionized, and by the finite time required for new
recombined and cooled gas to accrete in the Population B halos after the IGM
recombines again; but it is not clear if this would be sufficient to cause
the global instability in the star formation rate that would be required for
a double reionization.

  Finally, comparing our results with other recent work, we note that other
authors (e.g., Whyithe \& Loeb 2003; Haiman \& Holder 2003; Ciardi \etal
2003; Somerville \& Livio 2003; Sokasian \etal 2003; Cen 2003) have
generally emphasized that the central value of the {\it WMAP} result
$\tau_e\simeq 0.17$ can be obtained in CDM models with a physically allowed
emissivity of ionizing photons. Although we agree with this conclusion, we
note that a drastic change in the emissivity per unit mass of star-forming
halos from $z\gtrsim 15$ to $z=4$ is required. For example, Ciardi et
al. (2003) find, using our Model 3, that $\tau_e=0.16$ can be reached in
models where the dominant emission comes from halos corresponding roughly to
our Population A, but the emissivity they assume would vastly overpredict
the intensity of the ionizing background at $z=4$. At the same time, models
with no such change in the emissivity yield a lower optical depth that is
still within the plausible error range of the {\it WMAP} result. Our
approach is closest to that of Chiu et al. (2003), who use an ionizing
emission efficiency chosen to yield reasonable models for the ionizing
background, and our conclusions are generally in agreement with theirs.

\section{CONCLUSION}

  The first detection of the Thompson optical depth to the CMB by the {\it
WMAP} mission introduces a new era in the study of reionization.  The
observations of the Gunn-Peterson absorption in the highest redshift
quasars, which are sensitive to the time when reionization ended and the
mean free path of ionizing photons increased, can be confronted with the
total ionized column to the surface of last scattering, which is sensitive
to the epoch when about half of the baryons in the universe were
ionized. The simplest assumption we can make about the emissivity is that it
is roughly constant in all halos in which gas can cool. When we calibrate
the emissivity to the upper limit allowed by observations at $z=4$, we find
that the end of reionization is close to $z=6$, as suggested by
observations, but that the optical depth is smaller than $0.09$. To obtain
an optical depth as high as $\tau_e=0.17$ (the central value of the {\it
WMAP} result), an increase of the rate of ionizing emission per unit mass in
collapsed halos with redshift, of a factor $\sim 100$ to $300$ (depending on
the CDM power spectrum model) from $z=4$ to $z\sim 20$, is required.

  Further progress on our understanding of the reionization epoch will come,
among other things, from an increased accuracy in the measurement of
$\tau_e$, as well as more robust measurements of the power spectrum on small
scales from \lya forest observations. The discovery of quasars at $z>6$ and
the search for any small fraction of \lya transmitted flux in gaps along
their Gunn-Peterson trough (White \etal 2003) may also allow us to test if
the fraction of the ionized medium was close to unity for a wide redshift
range above $z=6$ (as in our models with high $\gamma$ in Fig.\ 3), or if it
increased rapidly over a short redshift range.  If an optical depth as high
as the central value of the {\it WMAP} measurement is confirmed, then either
a physical explanation for the high efficiency of the ionizing emissivity of
the first objects will need to be found, or some more fundamental
modification of the CDM model, such as non-Gaussian fluctuations on small
scales (e.g., Avelino \& Liddle 2003), may be implied.

\acknowledgments

  This work was supported in part by NSF grant NSF-0098515.


\clearpage

\begin{figure}
\plotone{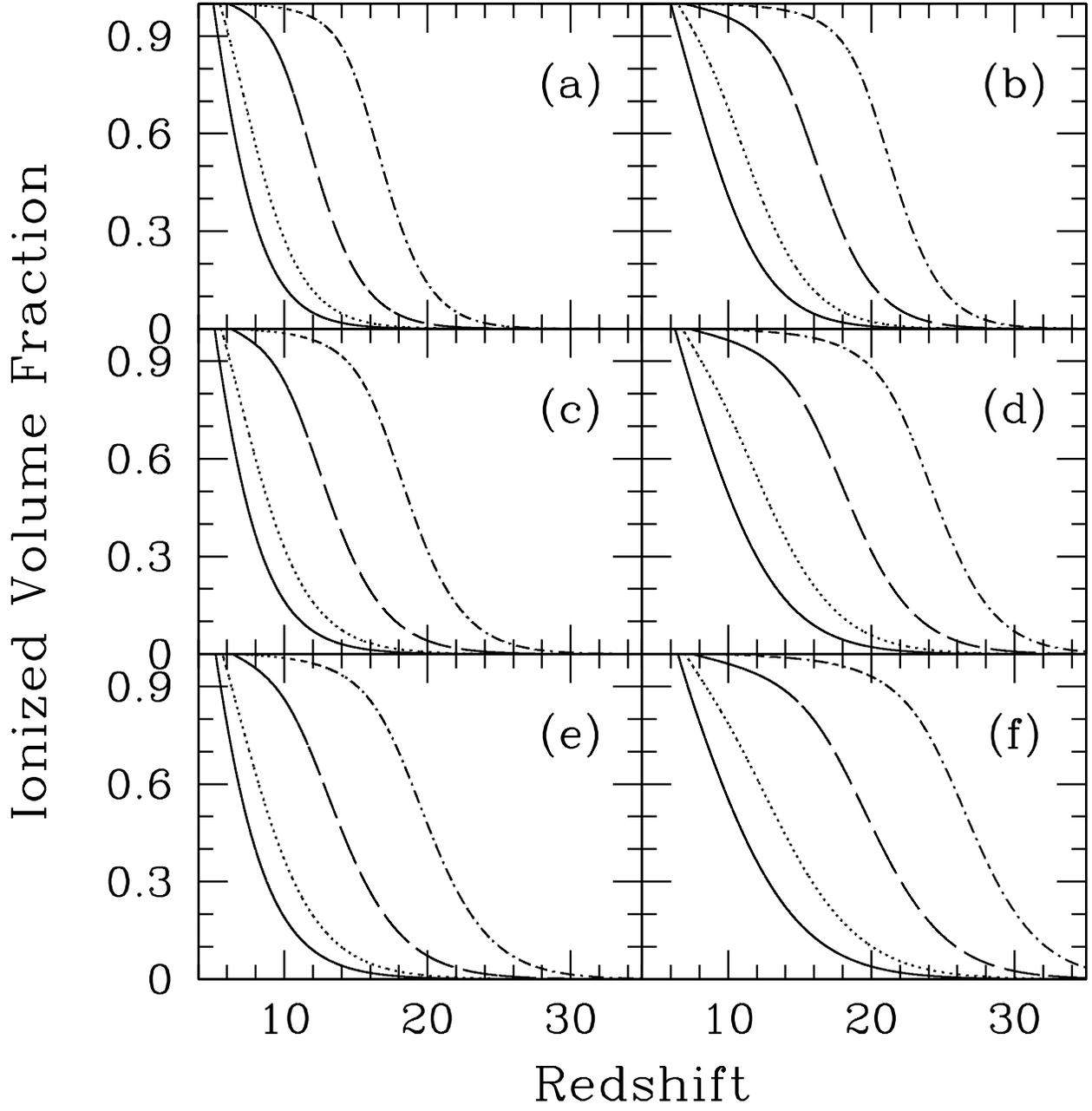}
\caption[f3.eps]{Reionization history of the universe for six cases: {\it
(a)} Model 1, $\alpha=0$; {\it (b)} Model 1, $\alpha=1.5$; {\it (c)} Model
2, $\alpha=0$; {\it (d)} Model 2, $\alpha=1.5$; {\it (e)} Model 3,
$\alpha=0$; {\it (f)} Model 3, $\alpha=1.5$. Within each panel, the lines
indicate the reionization history for $\gamma$ = 0 (solid), 1 (dotted), 10
(dashed), and 100 (dot-dash). \label{fig3}}
\end{figure}

\begin{figure}
\plotone{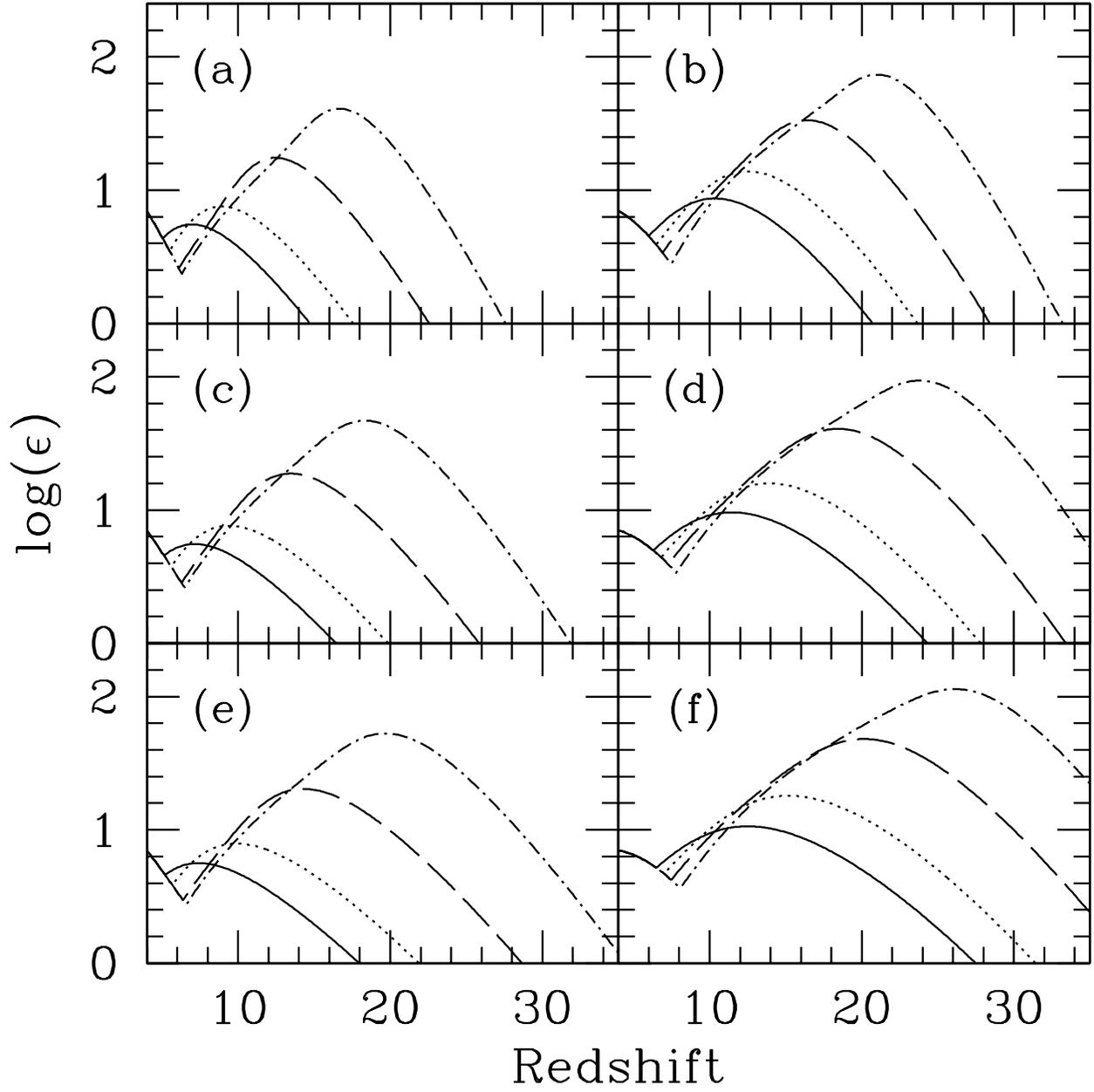}
\caption[f4.eps]{Log emissivity (ionizing photons per baryon per Hubble
Time at $z=4$) versus redshift.  Panels and lines as in Figure 3
\label{fig4}}
\end{figure}

\clearpage

\begin{deluxetable}{rcccccccc}
\tablecaption{Model Optical Depths \label{tab1}}
\tablewidth{0pt}
\tablehead{
\colhead{} & \multicolumn{2}{c}{Model 1} & \colhead{} & \multicolumn{2}{c}{Model 2} &
\colhead{} & \multicolumn{2}{c}{Model 3} \\
\cline{2-3}  \cline{5-6}  \cline{8-9}\\
\colhead{$\gamma$} & \colhead{$\tau_{e} (\alpha=0)$} & \colhead{$\tau_{e} (\alpha=1.5)$} &
\colhead{} & \colhead{$\tau_{e} (\alpha=0)$} & \colhead{$\tau_{e} (\alpha=1.5)$} &
\colhead{} & \colhead{$\tau_{e} (\alpha=0)$} & \colhead{$\tau_{e} (\alpha=1.5)$}
}
\startdata
0 & 0.047 & 0.071  & & 0.050 & 0.080 & & 0.053 & 0.088 \\
1 & 0.060 & 0.092  & & 0.064 & 0.106 & & 0.068 & 0.117 \\
10 & 0.098 & 0.146 & & 0.108 & 0.171 & & 0.117 & 0.194 \\
100 & 0.153 & 0.216 & & 0.176 & 0.259 & & 0.195 & 0.298 \\
\enddata
\end{deluxetable}

\end{document}